\documentclass[twocolumn]{emulateapj}

\def\mbf#1{\mbox{\boldmath ${#1}$}}
\def\Alfven{Alfv\'{e}n~}

\begin{document}
\title{Alfv\'{e}n Wave-Driven Supernova Explosion}
\author{Takeru K. Suzuki$^{1}$, Kohsuke Sumiyoshi$^{2}$, 
Shoichi Yamada$^{3,4}$}

\altaffiltext{1}{School of Arts \& Sciences, University of Tokyo, Komaba, 
Meguro, Tokyo, 153-8902, Japan}
\email{stakeru@ea.c.u-tokyo.ac.jp}
\altaffiltext{2}{Numazu College of Technology, Ooka 3600, Numazu, 
Shizuoka 410-8501, Japan}
\altaffiltext{3}{Science \& Engineering, Waseda University, Okubo, 
3-4-1, Shinjuku, 
Tokyo 169-8555, Japan}
\altaffiltext{4}{Advanced Research Institute 
for Science and Engineering, Waseda University, Okubo, 3-4-1, Shinjuku, 
Tokyo 169-8555, Japan}

\begin{abstract}
We investigate the role of \Alfven waves in the core-collapse supernova (SN) 
explosion. We assume that \Alfven waves are generated by convections inside 
a proto-neutron star (PNS) and emitted from its surface. Then these waves 
propagate outwards, dissipate via nonlinear processes, and heat up matter 
around a stalled prompt shock. To quantitatively assess the importance of this 
process for the revival of the stalled shock, we perform 1D time-dependent 
hydrodynamical simulations, taking into account the heating via the dissipation 
of \Alfven waves that propagate radially outwards along open flux tubes. 
We show that the shock revival occurs if the surface field 
strength is larger than $\sim 2\times 10^{15}$G and if the amplitude of velocity 
fluctuation at the PNS surface is larger than $\sim 20$\% of the local 
sound speed. Interestingly, the \Alfven wave mechanism is self-regulating 
in the sense that the explosion energy is not very sensitive to the surface field 
strength and initial amplitude of \Alfven waves as long as they are larger than the
threshold values given above. 
      
\end{abstract}
\keywords{MHD -- supernovae:general -- waves}

\section{Introduction}
The most promising scenario of collapse-driven supernova (SN) is currently supposed to be the 
so-called delayed explosion by neutrino heating \citep[][and references therein]{kot06,jan07}: the prompt shock, which was generated by
core bounce and stalled by neutrino emissions and dissociations of nuclei, is heated up and revived
by neutrinos coming out of the proto-neutron star (PNS), leading eventually to a SN explosion. 
Under spherical symmetry, however, no successful explosion has been obtained so far 
even though up-to-date micro-physics, such as equation of state 
and weak interaction rates, have been fully incorporated~\citep{jan07}.
Various other effects have also been explored over the years. 
The implications of stellar rotation and different sorts 
of hydrodynamical instabilities have been extensively 
studied.~(\citet{kot06} and see also \citet{mar07,mez07} for very recent progresses.) 

Magnetic fields are drawing much attention of researchers these days although the history of 
research is quite long. After some pioneering papers~\citep{leb70,bps76,mei76,sym84}, 
the subject had been forgotten for a while 
because it was realized that very strong magnetic fields are 
required to affect the supernova dynamics, which was supposed to be unrealistic at that time. 
The situation changed with the observational evidence of strongly magnetized neutron stars or magnetars 
\citep{tho96} and the progress in theoretical understanding of the magneto-rotational instability or MRI~\citep{bal91}.

The possible importance of MRI in the generation of magnetic field in core-collapse SNe 
was first pointed out by \citet{aki03}. 
A lot of papers have been published 
\citep{ys04,tqb05,mba06,kot06,bur07,wa07} to investigate possible roles of 
magnetic fields in the supernova dynamics, especially 
on their kinematical aspects such as magnetic pressure and
torque induced by rapid rotations in supernova cores. 
\citet{tqb05}, on the other hand, considered a MHD turbulence that is 
possibly induced by MRI as a source of viscosity to tap free energies stored in differential rotations. Although the rapid rotation of stellar core
is prerequisite in these studies, it may not be so easy to obtain in the presence of magnetic fields according to recent 
stellar evolution models because the transfer of angular momentum is 
efficient and the core may rotate rather slowly just prior to the 
gravitational collapse~\citep{hws05}\footnote{It is worth noting that 
\citet{bm07} recently pointed out a new mechanism to generate the rotation of PNSs. They claimed that the non-axisymmetric  
SASI might be a source of the angular momentum of PNSs. If true, no rotation may be needed prior to the collapse to 
account for the spin of neutron stars.}.

More recently, yet another supernova mechanism was put forward: sound waves  
generated by PNS oscillations of mainly g-mode nature, which are probably induced by turbulence caused by
the standing accretion shock instability (SASI), heat up matter through nonlinear dissipations, revive the stalled shock wave 
and produce explosions at very late times \citep{bur06}. Since this acoustic mechanism consists of several steps, whose efficiencies
appear to be not very high~\citep{yos07,mar07}, the viability of the mechanism is still controversial and further
explorations from various view points are needed. 

At present the supernova mechanism is still elusive in spite of these extensive efforts. 
The above-mentioned research trend naturally leads us to the exploration of still another type of waves that 
may also contribute to the supernova explosion: \Alfven waves. If the supernova core is magnetized, 
the oscillations of PNS will emit not only sound waves but also \Alfven waves. It is also possible that 
\Alfven waves are excited 
by convections in PNS, which have been demonstrated to exist after core 
bounce~\citep{kjm96}. 
This is quite analogous to what happens in the Sun \citep[e.g.][hereafter SI05; SI06]
{si05,si06}. A fraction of these waves propagate outwards and dissipate later through nonlinear processes, which then 
will heat up matter and may lead to the revival of the stalled shock.  Such a scenario as a supernova 
mechanism has not been studied on a quantitative basis so far,  
although it has been discussed in a qualitative manner \citep[e.g.][]{whe00,wj05,bur06} 
or in the context of the nucleosynthesis in $\nu$-driven winds \citep{sn05}.  
It should be emphasized that it is the dissipations of 
fluctuating components of magnetic fields associated with \Alfven waves that heat up matter and revive the shock wave, 
whereas in most of the previous papers, torques and pressures exerted by total magnetic fields were the
key players. Since the main aim in this paper is not to construct a realistic model but to elucidate the characteristics
of the \Alfven wave heating in a quantitative manner, we employ 1D simplified but dynamical simulations.     

\section{Models}

In this section, we describe the scenario we have in mind and explain 
nonlinear dissipation processes of \Alfven waves. 
Based on these we give the basic equations and employed approximations in 
detail.

\subsection{Basic Picture} 
Convections around the neutrinosphere that roughly coincides with the surface of a PNS are induced by 
negative gradients of lepton-fraction and entropy about a few tens of milliseconds 
after core bounce. According to hydrodynamical simulations of PNS 
convections by \citet{kjm96}, the amplitude of velocity fluctuations near the PNS surface 
is $\simeq 4\times 10^3$ km s$^{-1}$ on average and becomes as high as $10^4$ km s$^{-1}$. 
Recent simulations also reported that SASI in accreting flows
might excite a comparable order of surface fluctuations~\citep{oky06}. And not to mention, 
the non-radial oscillations of PNS are an obvious possibility as a source of \Alfven waves.   

These activities in PNS generate various modes of waves with the amplitude of velocity fluctuations mentioned 
above, which then emanate from its surface and propagate outwards. 
If the magnetic field is sufficiently strong, 
two types of magnetic waves are important: \Alfven wave and fast wave. 
The \Alfven waves propagate along field lines and are less subject to 
dissipations thanks to their incompressive character. On the other hand, the fast waves propagate almost isotropically and 
can traverse field lines although they suffer more damping owing to the compressive nature.  Therefore, 
\Alfven waves can transport energy farther away along magnetic filed lines unless they form closed loops. 
The fast waves could become important, however, in the equatorial region if PNS rotates rapidly and the 
field lines are tightly wound up. 
In this paper, we assume that open-field regions prevail in the post-bounce core SN core 
and study the energy deposition by the dissipations of \Alfven waves through nonlinear 
processes and its implications for the shock revival.

The energy source in this mechanism is eventually the gravitational energy that is released by the matter accretion onto
PNS. Some of the released gravitational energy is converted to the kinetic energy of 
surface fluctuations in the PNS by convections, SASI and other instabilities. Then the surface fluctuations excite 
various waves in PNS, some of which are \Alfven waves traveling outwards along field lines. This sequence of processes 
is quite similar to what Burrows et al. (2006) proposed in their acoustic mechanism. 
 

As will be shown later,  we assume strong magnetic fields of the magnetar scale, $\sim 10^{15}$G, on the PNS surface. 
We do not specify the origin of the strong magnetic fields in this paper (but see \S \ref{sec:mgf} for related discussions.) and
treat the field strength as a free parameter. As mentioned above, we focus on the propagation of \Alfven waves 
and pay attention only to magnetic flux tubes that are extended beyond the stalled shock    
because 
the \Alfven wave carry the energy parallel to a field line; if the magnetic 
field is closed near the PNS surface, \Alfven waves cannot transfer energy to 
the stalled shock.  

Since, as a first step, the aim of the paper is not to construct a realistic model, 
which should be the next step, but to understand the essential features of the \Alfven wave 
heating in the post bounce core, we make the models as simple as possible, incorporating only minimum ingredients. 
Since stellar rotation is not an indispensable player in the mechanism considered here, we just neglect it 
and assume radial magnetic fields as the simplest configuration for open field lines in this paper.  
The field strength is given by   
\begin{equation}
B_r = B_{r,0} \frac{r_{0}^2} { r^{2}}, 
\label{eq:nomon}
\end{equation}
where $B_{r,0}$ is the field strength at the PNS surface, $r=r_0$. 
We perform one-dimensional (1D) simulations, ignoring the effects of neighboring magnetic fields. 

Even if the stellar rotation is fast, our models will be still applicable locally to the polar 
region, where the centrifugal force is minimum, although the simple prescription for
the expansion factor of flux tubes, $\propto r^2$, adopted in this paper may need further 
elaboration. In the equatorial region, on the other hand, the validity of the assumption in this paper 
depends not only on the rotation period but also on the reconnection efficiency among 
the field lines frozen into the accreted matter. This issue will be discussed again in 
\S \ref{sec:mgf}.

\subsection{Dissipation of \Alfven Waves}
\label{dis:disaw}
The effects of the \Alfven waves on accreting matter are twofold: (1) the 
heating of matter by the \Alfven wave dissipations and 
(2) the extra pressure exerted by the \Alfven waves.
We take into account only the former in this paper.
The \Alfven wave with linear amplitude is non-dissipative 
due to the incompressive nature.  
However, nonlinear \Alfven waves suffer various dissipation processes. 
The nonlinearity, $w$, of \Alfven waves can be defined as 
\begin{equation}
w \equiv \frac{\delta v_{\perp}}{v_{\rm A}} = 
\frac{\delta B_{\perp}}{B_r},  
\label{eq:nnlnrt}
\end{equation}  
where $\delta v_{\perp}$ and $\delta B_{\perp}$ are the amplitudes of velocity and magnetic 
field, the subscript '$\perp$' denotes the transverse directions 
with respect to $B_r$, and $v_{\rm A}=B_r/\sqrt{4\pi\rho}$ is the \Alfven speed. 
Here we have used the relation, 
$\delta v_{\perp} = \delta B_{\perp}/\sqrt{4\pi \rho}$, that the \Alfven wave satisfies~\citep[\S 10 of ]
[]{lc99}.  
Even though the initial amplitude at the launch from the PNS surface is in 
the linear regime, $w \ll 1$, it grows thanks to 
the decrease of $B_r$ (Equation~(\ref{eq:nomon})). Eventually, the \Alfven 
wave becomes nonlinear, $w \sim 1$, and 
various dissipation processes set in. 

The excitation of compressive waves by nonlinear mode 
conversions is a route of the dissipation. 
If the \Alfven wave is not strictly circularly polarized, the 
magnetic pressure, $\delta B_{\perp}^2/8\pi$, fluctuates along with 
$B_r$ and induces longitudinal compressive 
motions, most of which correspond to slow magnetohydrodynamical (MHD) waves 
(SI05;SI06). Even if the \Alfven wave is circularly polarized, it is subject to the 
parametric decay instability, which generates outgoing slow MHD waves 
and incoming \Alfven waves \citep{gol78,ter86}. The velocity amplitudes 
of the slow waves are also amplified as they propagate outwards and the density 
decreases. Eventually the wave fronts steepen to form shocks and heat up matter. 

If $B_r$ has a transverse gradient (along $\perp$ direction), fast MHD waves 
that propagate perpendicularly to $B_r$ are also excited from \Alfven 
waves \citep{nrm97}. These fast waves also heat ambient matter by the shock formation. 
Moreover, the transverse inhomogeneity of $B_r$ leads to the phase mixing \citep{hp83}, 
which is another channel of the dissipation of \Alfven waves.

The turbulent cascade may also work in the dissipation of \Alfven waves. 
In the PNS, the \Alfven speed is not constant along the radial 
magnetic field because of the changes in both $B_r$ and density. 
Then, incoming \Alfven waves are excited from the outgoing ones by the deformation of wave shapes
and the interactions between the outgoing and incoming \Alfven waves lead to the formation of 
smaller scale (large wave number) structures mainly in the perpendicular 
directions \citep{gs95}. This turbulent cascade to higher wave number proceeds up to the dissipation 
range, where resistivities (for magnetic field fluctuations) and/or viscosities (for velocity fluctuations) 
become important. Then, the energy that \Alfven waves are carrying is finally transferred to the 
ambient matter. 

As a result of these various dissipation processes, the nonlinearity, $w$, of 
the outgoing \Alfven waves is saturated at a certain level. 
According to dynamical simulations of solar (SI05;SI06) and stellar 
\citep{suz07} winds, the saturation level is found to be $w \lesssim 0.3-1$, which we apply in this paper  
to the \Alfven waves in the supernova core. It should be noted that the saturation with a constant $w$ 
implies that $\delta B_{\perp}$ itself decreases as the \Alfven wave propagate further outwards, since 
$B_r$ is declining (see Equation~(\ref{eq:nnlnrt})). This means the dissipation of the energy of \Alfven waves 
and results in the heating of the matter. 

\subsection{Formulation}
We evaluate the \Alfven wave heating  by considering the conservation of wave energy 
under the WKB, or short wave length, approximation.  
This treatment enables us to obtain the estimations in a 
simple manner without solving the fully nonlinear MHD 
equations. 
The conservation of wave energy can be expressed \citep{jaq77} as 
\begin{equation}
\frac{\partial {\cal{E}}_{\rm w}}{\partial t} +\mbf{\nabla \cdot F_{\rm w}}
- \mbf{v \cdot \nabla} P_{\rm w} = - \rho \dot{q}_{\rm w} ,
\label{eq:gnwv}
\end{equation} 
where $\rho, \mbf{v}$ are the density and velocity of the accretion flow and 
${\cal{E}}_{\rm w}$
is the wave energy density, 
$\mbf{F_{\rm w}}$ is the wave energy 
flux, $P_{\rm w}$ is the wave pressure, and $\dot{q}_{\rm w}$ is the dissipation rate of wave energy per unit mass. As for the \Alfven wave, 
${\cal{E}}_{\rm w}=\rho \langle \delta v_{\perp}^2\rangle 
= \langle \delta B_{\perp}^2\rangle/4\pi$ and $P_{\rm w}={\cal{E}}_{\rm w}/2$, 
where 
the bracket $\langle \cdots \rangle$ stands for the average 
over a period of the \Alfven wave. 
The last term on the left hand side denotes the rate of work done by the 
\Alfven waves on the accretion flow, which we neglect in this paper as 
stated above. 
The right hand side represents the energy deposition by the \Alfven waves 
to the accreting matter. 
When $\dot{q}_{\rm w}>0$,  the matter is heated up by the wave 
dissipations, whereas the \Alfven waves travel without dissipation if $\dot{q}_{\rm w} =0$. 

It is convenient for later use to introduce an adiabatic constant, the 
so-called wave action, $\mbf{H_w}$, defined as follows \citep{jaq77}:   
\begin{equation}
\mbf{\nabla \cdot F_{\rm w}}  
-v_r \frac{dP_{\rm w}}{dr}   
\equiv \frac{v_{\rm A}}{v_{\rm A} + v} \mbf{\nabla \cdot H_{\rm w}} .   
\label{eq:wveq}
\end{equation}
We neglect relativistic corrections 
because they are minor outside the PNS. It should be noted that the 
wave action, $\mbf{H_{\rm w}}$, instead of $\mbf{F_{\rm w}}$, is conserved in moving media. 
The specific form of $H_{\rm w}$ is given as \citep{jaq77}
\begin{equation}
\mbf{H_{\rm w}} = \frac{\langle \delta B_{\perp}^2\rangle}{4\pi}
\frac{(v_{\rm A}+v)(\mbf{v_{\rm A}+v})} {v_{\rm A}} .
\end{equation}
In this paper we assume the steady propagations of \Alfven waves and 
neglect the time derivative, $\frac{\partial}{\partial t}$, in Equation~(\ref{eq:gnwv}).
This is valid when the \Alfven transit time is shorter than a typical time-scale
of the system, which will be discussed in \S\ref{sec:tw}.

The initial amplitude of \Alfven waves at the PNS surface is supposed to 
be an order of convective velocities at the surface of PNS, which are suggested to be 
a fraction of the sound speed\footnote{This is very similar to the surface convection 
in the Sun. The observed granulation speed is $1-2$ km s$^{-1}$, while the sound speed 
at the photosphere is $\approx 5$ km s$^{-1}$.}: 
\begin{equation}
\delta v_{\perp,0} = \epsilon c_{\rm s,0}, 
\end{equation} 
where $c_{\rm s,0}$ is the sound speed at the PNS surface and 
we choose $\epsilon = 0.1 - 0.3$ in our simulations. 
Because $c_{\rm s,0}$ is $\approx$10\% of the light speed, 
$\epsilon = 0.1 - 0.3$ corresponds to $\delta v_{\perp,0}=(0.3-1)\times 
10^4$ km s$^{-1}$, which is comparable to the values obtained in 
hydrodynamical simulations \citep{kjm96}. 
For a sufficiently large background magnetic field at the PNS surface, 
$B_{r,0}\gtrsim 5\times 10^{14}$ G, the initial wave amplitude is small in
the sense that $w < 0.1$.
In such a condition, the \Alfven wave
travels outwards without dissipation ($\dot{q}_w = 0$) near the PNS surface 
and its amplitude evolves according to the following relation: 
\begin{equation}
\frac{\langle \delta B_{\perp}^2\rangle}{4\pi}
\frac{(v_{\rm A}+v_r)^2} {v_{\rm A}} r^2 = H_{\rm w,0} r_0^2, 
\end{equation}
where 
$H_{\rm w,0}$ is the wave action at its surface, 
and $v_r$ is the radial velocity of the background accretion flow. 

As the \Alfven waves travel outwards, 
the nonlinearity, $w$, increases because 
of the expansion of the radial magnetic flux tube (Equation~(\ref{eq:nomon})).
The dissipation of \Alfven waves eventually sets in when they reach the 
nonlinear regime as discussed above   
and the nonlinearity of the \Alfven waves is saturated at a certain level: 
\begin{equation}
w \approx \alpha. 
\label{eq:nonli}
\end{equation}    
In this paper we adopt a constant $\alpha=0.5$ as a standard saturation 
level, based on our previous results on the solar and stellar winds 
(SI05;SI06;Suzuki 2007), which showed that $w (\lesssim 1)$ is more or less constant 
or very slowly varying as a function of $r$ \citep[see also e.g.][for steady-state modeling]{hol73}. 
This simple prescription of the constant saturation level is expected to incorporate phenomenologically 
all the complex physical processes of \Alfven wave dissipations discussed in 
\S\ref{dis:disaw} and provide us with a reasonable heating rate.  


For the steady state, the energy dissipation rate, $\dot{q}_w$, 
can be obtained from the conservation of the wave action without referring to the details of 
the nonlinear processes that are responsible for the dissipations.
Using the relation in the dissipation region, $\delta B_{\perp} \propto B_r 
\propto (r_0/r)^2$ (Equation~(\ref{eq:nomon})), the heating rate, 
$\dot{q}_w$ (erg g$^{-1}$s$^{-1}$), 
can be expressed as
\begin{equation}
\dot{q}_{\rm w} = \frac{v_{\rm A}}{v_{\rm A}+v_r}\frac{\alpha^2 B_r}{4\pi \rho}
\frac{d}{dr}((v_{\rm A}+v_r)^2\sqrt{4\pi \rho}). 
\label{eq:heating}
\end{equation}
An important point here is that we can evaluate $\dot{q}_{\rm w}$ from the local 
distributions of $\rho$, $v_{r}$, and $B_{r}$. 

\subsection{Simulations}

Incorporating the above formula for the heating by \Alfven waves, we perform 1D 
time-dependent simulations of the post-bounce evolutions of SN core. 
We use the 15$M_{\odot}$ progenitor star of \citet{ww95} as an initial 
condition, and employ a numerical code~\citep{sum05} to solve 
general relativistic hydrodynamics and neutrino transport, adding 
the extra heating term given by Equation~(\ref{eq:heating}) in the energy equation.     
The \Alfven wave heating ($\dot{q}_{\rm w}$) is switched on at $100$ ms
after core bounce. We have in mind that this is the time for the development 
of  (magneto-)convection that drives \Alfven waves \citep{kjm96,aki03,mst06}.  
Incidentally, we define the PNS surface as the position of the density, 
$\rho_0 =10^{11}$ g cm$^{-3}$, which, as a consequence, becomes smaller as the PNS contracts. 

We have three parameters, $B_{r,0}$, $\epsilon$, and 
$\alpha$ in the above prescription. We fix the saturation level, $\alpha=0.5$, 
which controls the location of wave dissipations. The other two, the field strength, $B_{r,0}$, 
at the PNS surface and the normalized initial amplitude, $\epsilon$, of perturbations, determine 
the wave energy injected from the PNS surface. In the following, we explore the condition on 
these two parameters for the shock revival by simulating nine models 
in Table \ref{tab:sum}.

\section{Results}
\label{sec:res}
\begin{deluxetable}{ccccccc}
\small
\tablecolumns{7}
\tablecaption{Summary of Simulations.\label{tab:sum}}
\tablehead{
\colhead{Model} & \colhead{$B_{r,0}$(G)} & \colhead{$\epsilon$} 
& \colhead{Explosion} & \colhead{$E_{\rm exp}$} 
& \colhead{$M_{\rm ej}$} 
& \colhead{$M_{\rm cut}$} }
\startdata
I & $1\times 10^{15}$ & 0.1 & No & --- & --- & --- \\
II & $1\times 10^{15}$ & 0.2 & No & --- & --- & --- \\
III & $1\times 10^{15}$ & 0.3 & No & --- & --- & --- \\
IV & $2\times 10^{15}$ & 0.1 & Marginal & --- & --- & --- \\
V & $2\times 10^{15}$ & 0.2 & Yes & $1.2$ & 0.08 & 1.38\\
VI & $2\times 10^{15}$ & 0.3 & Yes & $1.6$ & 0.10 & 1.37\\
VII & $3\times 10^{15}$ & 0.1 & Yes & $0.33$ & 0.04 &1.41\\
VIII & $3\times 10^{15}$ & 0.2 & Yes & $1.5$ & 0.10 & 1.38 \\
IX & $3\times 10^{15}$ & 0.3 & Yes & $2.2$ & 0.14 & 1.38
\enddata

\tablecomments{The explosion energy, ejecta mass and 
PNS mass are denoted by $E_{\rm exp}$, $M_{\rm ej}$ and $M_{\rm cut}$, 
respectively. The unit of $E_{\rm exp}$ is $10^{51}$erg and the units of 
$M_{\rm ej}$ and $M_{\rm cut}$ are $M_{\odot}$.}
\end{deluxetable}

\begin{figure}[t]
\epsscale{0.65} 
\plotone{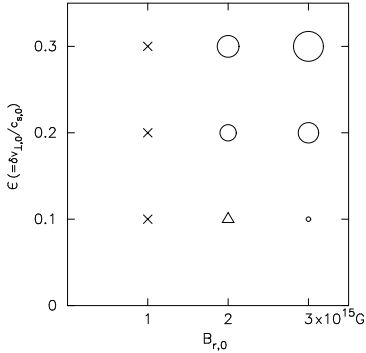} 
\caption{Status of each model. 
The circles and crosses respectively corresponds to the models that result 
in shock revival and the models that do not give shock revival. 
The triangle is a marginal case. 
The radii of the circles in the successful cases are scaled by $E_{\rm exp}$.}
\label{fig:sum}
\end{figure}

Figure \ref{fig:sum} displays
the success or failure of shock revival in the plane of the field strength at the 
PNS surface, $B_{r, 0}$, and the initial amplitude of velocity perturbation, $\epsilon$. 
Table \ref{fig:sum} gives more detailed information on the ejected mass, $M_{\rm ej}$, 
explosion energy, $E_{\rm exp}$, and remnant ($\approx$ PNS) mass, $M_{\rm cut}$, 
for the models that obtain the shock revival under the current approximation. They are estimated at the time $t=200$~ms 
after core bounce as follows. We first define the ejecta as the collection of mass shells with positive 
total energy, which is the sum of kinetic, internal and gravitational energies. Then $M_{\rm ej}$ and 
$E_{\rm exp}$ are obtained as the sums of mass and total energy, respectively, of 
each mass shell that comprises the ejecta. In so doing, the non-relativistic expression
is employed for the energy.
It is clear that an explosion with $E_{\rm exp}\ge 10^{51}$ erg is obtained if 
the magnetic field at the PNS surface is strong, $B_{r,0}\gtrsim 2\times 10^{15}$G, 
and if the initial amplitude of \Alfven waves is larger than a certain value, 
$\epsilon(=\delta v_{\perp}/c_{\rm s})\gtrsim 0.2$. Note that the surface field strength
is of the same order as those inferred for magnetars.

\begin{figure}[b]
\epsscale{0.9} 
\plotone{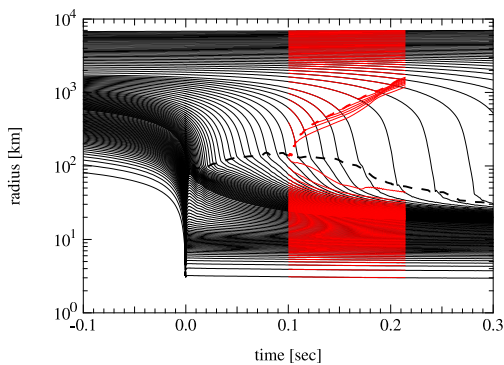} 
\caption{Time evolutions of mass shells. Model V with \Alfven wave heating 
($B_{r,0} = 2\times 10^{15}$G \& $\epsilon = 0.2$; red lines) is superimposed on
the original non-magnetized model (black lines). The time is measured from 
the core bounce. The dashed lines show the locations of shock front.}
\label{fig:str}
\end{figure}

Figure \ref{fig:str} presents a typical \Alfven wave-driven shock revival, in which 
the result for model V (red lines) is superimposed on the original result 
for the non-magnetized spherically symmetric model (black lines) \citep{sum05}. 
The trajectories of mass shells are plotted against the time from core bounce. 
The figure clearly demonstrates the shock revival by the \Alfven wave heating  
for the otherwise failed neutrino heating model \citep{sum05} like those discussed 
in many previous papers \citep[e.g.][and references therein]{kot06}.


\begin{figure}
\epsscale{0.9} 
\plotone{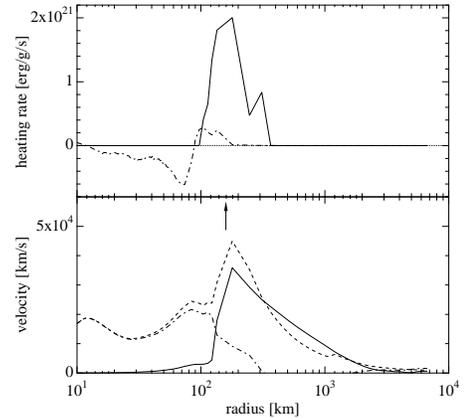} 
\caption{{\it Top}: The heating (cooling for negative values) rates from 
neutrino (dot-dashed) and \Alfven waves (solid)
of model V at $t=100$ms. The arrow indicates the location of the shock front. 
{\it Bottom}: The distributions of $-v_r(>0)$ (solid), $v_{\rm A}$ (dashed), and $v_r+v_{\rm A}$ (dot-dashed) 
at $t=100$ms for model V.}
\label{fig:heat}
\end{figure}

Figure \ref{fig:heat} shows the rates of \Alfven wave heating and 
neutrino heating (top panel) along with the velocity distribution (bottom panel) at $t=100$ ms. 
The top panel demonstrates that the \Alfven wave heating operates mainly in the 
vicinity of the stalled shock wave and dominates the neutrino heating. 
The main reason for the localization of the \Alfven wave heating in the
{\it Eulearian} frame is the trapping of \Alfven waves. 
The propagation speed of the outgoing \Alfven wave is $v_{\rm A}+v_r$ in this frame. 
It rapidly decreases from $2\times 10^4$ km s$^{-1}$ at $r=100$ km 
to $\approx 0$ km s$^{-1}$ at $r=300$ km for model V (bottom panel of 
Figure \ref{fig:heat}), for example. 
The \Alfven waves cannot travel further outwards and are trapped inside 
$r \lesssim 300$ km in model V, so that they spend a long time there to 
damp almost completely. In the {\it Lagrangian} frame that moves at the inflow
velocity of the accreting matter, on the other hand, the \Alfven waves propagate 
outwards at the speed $v_{\rm A}$. This means that the inflowing matter is heated up 
by the dissipation of \Alfven waves just when it reaches the vicinity of the stalled shock wave.    


The luminosity, $L_{\rm A}$, of \Alfven waves at the PNS surface can be 
estimated under the assumption of spherical symmetry as follows:  
\begin{eqnarray}
L_{\rm A} &=&\frac{1}{2}\rho_0 \delta v_{\perp,0}^2 v_{\rm A,0} 
4 \pi r_0^2 
\nonumber \\
&\approx& 10^{52}{\rm erg\; s^{-1}}
\left(\frac{\rho_0}{10^{11}{\rm g\; cm^{-3}}}\right)^{1/2}
\left(\frac{c_{\rm s,0}}{0.1c}\right)^{2}\left(\frac{\epsilon}{0.2}\right)^2
\nonumber \\
& &\hspace{3cm}\left(\frac{B_{r,0}}{2\times 10^{15}{\rm G}}\right)
\left(\frac{r_0}{50{\rm km}}\right)^2. 
\end{eqnarray}
This implies that the emission of \Alfven waves  
for $\sim 100$ ms gives the energy injection of $\sim 10^{51}$erg. 
In most cases, almost all the energy of \Alfven waves is absorbed thanks to the trapping
of \Alfven waves just mentioned. This is also confirmed by the comparison with the 
simulation results. We can calculate the total heating rate (erg g$^{-1}$) by multiplying the heating 
rate per unit mass (erg g$^{-1}$s$^{-1}$; the top panel of Figure~\ref{fig:heat}) 
by the density ($\approx 10^9$g cm$^{-3}$) and volume of the heating region, 
$4\pi r^2 \Delta r$, where $\Delta r (\approx 100{\rm km \;\; in \;\; 
Figure~\ref{fig:heat}})$ is the thickness of the heating region.  
We thus obtain the total heating rate of $\approx 10^{52}$erg s$^{-1}$ for model V, which indicates that the injected 
\Alfven wave luminosity is mostly used for heating up the stalled shock.   

The regions with fast accretion velocities are 
preferentially heated up and eventually start to move outwards, provided $L_{\rm A}$ is 
sufficiently large. Once the stagnated shock wave is re-launched, the heating 
is reduced because \Alfven waves become untrapped again. 
For larger $L_{\rm A}$ the shock revival occurs earlier and the duration of heating, 
$\Delta \tau_{\rm A}$, is shorter.   
As a result, $E_{\rm exp}$ (roughly $\propto L_{\rm A}\Delta \tau_{\rm A}$) 
is not very sensitive to $B_{r,0}$ and $\epsilon$. 
In fact, although $L_{\rm A}$ of model IX is larger than that of model V by 
more than a factor of 3, the difference in $E_{\rm exp}$ is less than a factor 
of 2. In this sense the \Alfven wave mechanism is self-regulating. 
Interestingly, the acoustic wave mechanism is also claimed to be self-regulating 
\citep{bur06} though the regulating mechanism is different:  
the generation of the acoustic waves continues until the shock is revived and matter  
ceases to accrete.   

Model VII is exceptional among the explosion cases, giving a very 
small explosion energy. The \Alfven wave heating operates in a much outer region 
in this case because \Alfven waves become nonlinear, $\delta B_{\perp}/B_r>\alpha$,
only after crossing the shock wave owing to the large $B_{r,0}$ and small $\epsilon$. 
A sizable fraction of $L_{\rm A}$ leaks out of the stalled shock wave and
a tiny amount 
is ejected with a quite small $E_{\rm exp}$.
 
The models with $B_{r,0} = 1\times 10^{15}$ G produce no explosion.   
This is first because $L_{\rm A}$ itself is small (mainly models I \& II) 
owing to the small $B_{r,0}$ 
and second 
because the dissipation of \Alfven waves occurs too early (mainly model III).
Note in particular that $L_{\rm A}$ of model III is larger than those of the explosion 
cases V and VII. 
In this case \Alfven waves dissipate in the inner region and the temperature
is increased there. As a result, 
the energy deposited by \Alfven waves is mostly 
converted to neutrino emission in this case. 


\section{Discussion}
\label{sec:dis}

In this section, we discuss more in detail the validity of the assumptions and approximations employed 
for the background magnetic field and the formulation of \Alfven wave propagations in 
this paper.

\subsection{Magnetic Field}
\label{sec:mgf}
Since we have seen that strong magnetic fields ($B_{r,0}\gtrsim 2\times 10^{15}$ G), which
are of the magnetar scale and much larger than those for the ordinary radio pulsars, 
are necessary to revive the stalled prompt shock by the \Alfven wave heating,  
one may think that the mechanism considered in this paper is only applicable to 
magnetar-forming supernovae and the \Alfven wave is not a major ingredient in 
the ordinary SN explosion. We cautiously note, however, that this strong magnetic field is required not for the 
ordinary neutron stars as we observe them but for the PNSs in their very infancy. 
As a matter of fact, there is a speculation that the 
magnetic fields of nascent PNS may be temporarily very strong and then decrease 
to the `normal' value ($\sim 10^{12}$ G) by energy releases occurring 
during the SN explosion and later evolution \citep{whe02}. 
If this is true, the \Alfven wave mechanism may work in a larger population of core-collapse SNe. 
  
The origin of such strong magnetic fields is still controversial.     
One possibility is referred to as the fossil origin hypothesis: the strong magnetic field in compact stars
is simply a consequence of the compression of the magnetic field that already exists in 
OB progenitors prior to the gravitational collapse. In fact, several magnetic massive stars have been observed to 
have a magnetic field whose average dipole-field strength is $\sim 1000$G \citep{nei03,hub06,don06}. 
The total magnetic flux of these stars is comparable to that of a typical magnetar \citep{fw06}, which 
implies that an additional generation and/or amplification of magnetic fields will not be necessary to 
obtain a highly magnetized compact remnant for these stars. 

Another possibility is an amplification of weak magnetic fields by the 
MRI \citep{aki03,mst06}, 
in which stellar rotation plays a key role, winding poloidal fields and driving the instability.  
In this scenario, toroidal magnetic fields are efficiently produced, whereas in the case of the fossil origin, 
we expect the radial component of magnetic field is dominant.   

Since the \Alfven wave carries energy along a field line, we are interested only in open magnetic flux tubes
that are extend beyond the stalled shock wave. As the simplest configuration, we have considered radial magnetic fields 
and neglected the toroidal component in this paper. As mentioned above, the approximation is justified if 
the magnetic field is of fossil origin and the progenitor is a slow rotator. 
Even if the progenitor core is a rapid rotator, our models will be still applicable to the polar region, where the 
effect of rotation is not strong and the toroidal magnetic fields are less important, although the radial 
dependence of the field strength may need elaboration. 
 
In the equatorial region of a rapidly rotating supernova core, on the other hand, the situation is much more  
complicated. Field lines are not directed radially in general and some of them may be closed, as expected for 
the dipole configuration. In addition, the continuous downward advection of magnetic fields may cause 
reconnections in the PNS, which in turn will open up some field lines again. In any case the toroidal component
will be dominant over the radial component \citep{bur07} and 
we need to include the effects of these spiral magnetic fields as well as rotation itself in discussing
the \Alfven wave heating in supernova cores quantitatively, which will be the future task.

\subsection{\Alfven Wave Propagation}
\label{sec:tw}
The treatment of \Alfven waves in this paper is admittedly a crude approximation. We employ the  
non-relativistic, steady-state, and WKB approximations for describing the propagation of 
\Alfven waves. Among the assumptions, the non-relativity is adequate for the \Alfven 
waves launched from the PNS surface since the relativistic corrections are indeed 
minor outside the PNS. 
The steady-state approximation is also reasonable because the \Alfven transit 
time is shorter than the expansion time of 
accreting matter; while the expansion time-scale of the ejecta is 
$\sim 50-100$ ms, the time for the \Alfven wave to travel from the PNS 
surface ($r\approx 50$ km) to the wave trapping region around the stalled 
shock ($r\approx 200$ km) is $\sim 10$ ms for $v_{\rm A}\approx 2\times 10^4$ 
km s$^{-1}$.

The WKB approximation is acceptable if the wavelength is shorter than the 
scale height of the background. The typical period of \Alfven waves generated in the PNS 
is supposed to be $\tau \sim 1$ ms, corresponding to the dynamical time-scale. Then, 
the wavelength becomes, $\lambda \approx v_{\rm A} \tau \sim 
2\times 10^4 \;{\rm km\; s^{-1}}\;\times \; 10^{-3} \;{\rm s}\sim 20$ km. Since the density 
scale height is shortest near the PNS surface and is $H_\rho\approx 30$km, 
\Alfven waves might be partially reflected there due to the deformation of the wave shape.   
In more detailed study, this effect should be taken into account together with the wave
pressure ignored in this paper.


\section{Conclusion}
In this paper we have studied the matter heating by \Alfven waves in the post-bounce supernova core
 and its implications for the shock revival. In order to elucidate the essential features 
of the mechanism quantitatively, we have done a couple of 1D dynamical simulations, 
neglecting rotation and toroidal magnetic fields but employing the 
Alfv\'{e}n-wave heating rate
based on our model of the nonlinear damping of \Alfven waves. 

We have found that if the surface magnetic field strength is $\gtrsim 2\times 10^{15}$G and 
if the surface velocity fluctuation is $\epsilon \gtrsim 0.2$, which corresponds to 
$\delta v_{\perp,0}\gtrsim 6\times 10^3$km s$^{-1}$, the stalled shock is revived by the 
\Alfven wave heating with a canonical explosion energy, $E_{\rm exp}\gtrsim 10^{51}$erg. 
The current mechanism is self-regulating in the sense that the explosion energy is not 
very sensitive to the surface field  strength and initial velocity fluctuation as long as 
they satisfy the above conditions. 
The above strong magnetic field is not a requirement for the ordinary  
neutron stars as observed but for the PNSs in their infancy. If magnetic fields decay through 
their subsequent evolution (Wheeler et al.2002), the \Alfven waves may play an important role 
in a larger population of core-collapse SNe.  

It has been also found that the wave trapping is essential in localizing 
the \Alfven wave heating in the vicinity of the stalled shock wave as well as in
regulating the explosion energy. In fact, if the magnetic field is weaker 
($\lesssim 2\times 10^{15}$ G), 
the \Alfven wave heating takes place much closer to the PNS because the \Alfven wave
becomes nonlinear earlier on at smaller radii. 
Then no shock revival occurs because the dissipated energy is 
mostly lost by neutrino cooling; neutrino emissions are enhanced in this 
case.  If the initial velocity fluctuation is smaller 
($\epsilon \lesssim 0.2$) with a stronger magnetic field, $\gtrsim 3\times 10^{15}$G, on the other hand,
most of \Alfven waves propagate through the stalled shock region with only a small amount of energy 
being deposited, which then results in a weak explosion with $E_{\rm exp} < 10^{51}$erg. 
It is noted, however, that even in these cases the \Alfven wave heating will 
be still important in supplementing the neutrino heating. 

As a first step, the models presented in this paper have much room for sophistication. 
Among other things, as mentioned repeatedly, rotation and toroidal magnetic fields should be 
somehow taken into account. 
In reality, the \Alfven wave mechanism probably works in cooperation with 
these processes. 
This will require multi-dimensional numerical modellings and will be 
the future work.

\acknowledgements
TKS thanks Shigehiro Nagataki and Toshi Ebisuzaki for many fruitful 
discussions. The authors thanks Hidetomo Sawai for references of magnetic 
OB stars.   
We also thank the second referee for valuable comments. 
This work was supported in part by Grants-in-Aid for Scientific Research
from the Ministry of Education, Science and Culture of Japan　(TKS : 19015004, 
KS : 18540291, 18540295, 19540252, SY : 14079202 and 14740166),　Inamori 
Foundation (TKS), and a Grant-in-Aid for the 21st century COE
program　``Holistic Research and Education Center for Physics of　
Self-organizing Systems''(SY).
The numerical simulations were performed
at NAO (iks13a, uks06a), JAEA and YITP.

\end{document}